\documentclass[12pt, twocolumn]{article}
\usepackage{amsmath,latexsym}
\usepackage{graphicx}
\begin{document}

 \title{\Huge Wormholes supported by Scalar Fields with Negative Kinetic Energy
  }
 \author{F.Rahaman$^*$, M.Kalam$^{\dag}$
  and K. A. Rahman$^*$}

\date{}
 \maketitle
 \begin{abstract}
We provide a  new model of  higher dimensional wormholes supported
by phantom energy derived from a scalar field with negative
kinetic term. We have shown that the Averaged Null Energy
Condition ( ANEC )
                  violating phantom energy can be reduced as
                  desired. We have also noticed that dimension of
                  the spacetime plays a crucial role for measuring
                  the ANEC
                  violating matter needed.

\end{abstract}

  \footnotetext{

 $*$Dept.of Mathematics, Jadavpur University, Kolkata-700 032, India

                                  E-Mail:farook\_rahaman@yahoo.com\\
$\dag$Dept. of Phys. , Netaji Nagar College for Women, Regent Estate, Kolkata-700092, India.\\
E-Mail:mehedikalam@yahoo.co.in\\
}
    \mbox{} \hspace{.2in}

\title{ \underline{\textbf{ I. Introduction:}} }

During last five or six years, scientists around the world follow
more or less the same path to explain the accelerating phase of
the Universe and to describe the wormhole physics. The first one
is an experimental evidence observed first by two group of
scientists, Perlmutter et al [1] and Riess et al [2] about 10
years ago. Whereas the second phenomena is a purely theoretical
prediction, predicted by Morris and Thorne [3] about 20 years
back. Almost all the authors believe that this acceleration is due
to the cause of unknown dark energy. The exciting and unusual
geometry ( hypothetical short cut in spacetime ) that aries as the
solutions of Einstein's equations is the wormhole. The most
disappointing fact that it needs unusual matter or energy ( termed
as exotic matter ). But fortunately, the spheres of action of
these heavenly sources ( dark energy and exotic matter ) are same
i.e. to produce negative pressure.  So wormhole physicists and
cosmologists exchange their ideas to achieve their goals.
Cosmologists have no headache , what amount of dark  energy to be
needed to fulfill their target. But wormhole physicists should
have to keep in view the minimizing the total amount of exotic
matter. Most of the works done considering the ideas of phantom
energy characterized by the equation of state ( EOS ) $p = w
\rho$, with $w<-1$, where,  $p$ and $\rho$ are respectively, the
pressure and energy density of the source and Chaplygin gas EOS
characterized by $p = a\rho - \frac{b}{\rho^\alpha}$ , ( $a$,
$b$  and $\alpha$ are positive constants  ) in their respective
fields of interest ( i.e. explaining the accelerating state of
the Universe and discussing wormhole physics ) [4-20].

\pagebreak

Recently, some cosmologists have considered phantom dark energy
models with negative kinetic term to explain the accelerating
phase of the universe[21-23].\\
 In this
work, we have borrowed from cosmologists the idea of phantom
energy that is generated by a scalar field Lagrangian with a
negative kinetic term.  We have developed the higher dimensional
wormhole geometry using the above phantom source. Due to the
string theory the gravity is a truly higher dimensional
interaction which becomes effectively four dimension at low
energies. Newtonian theory can not be obtained as a limit of
Einstein's theory. Wormhole structure can not be obtained from
Newtonian theory. But it is argued that wormhole like geometry
could be found in Einstein's theory. Moreover, to unify gravity
with other fundamental forces in nature, many theories such as
Super string theories, M theories etc demand extra spatial
dimension to be consistent. For this reason, why we consider
higher dimension to discuss wormhole structure. The plan of the
article is as follows: In the second section, we have presented
the basic equations for modelling the wormhole spacetime. In
section three, we have obtained the solutions of the field
equations. The properties of the wormholes are discussed in
section four.  Last section is devoted to a brief summary and
discussions including the calculations of the total amount of
average null energy condition violating exotic matter needed.
\\
\\
\\
\\
\\
\\
\\
\\
\\
\\
\title{ \underline{\textbf{ II. Construction of wormholes:}} }

For the  present study the metric for static spherically symmetric
spacetime in higher dimension is taken as
\begin{equation}
               ds^2=  - e^{2\alpha(r)} dt^2+ e^{2\beta(r)} dr^2+r^2 d\Omega_d^2
         \label{Eq3}
          \end{equation}
\\
\\
 \\
where, The line element ${d{\Omega}_d}^2$ on the unit $d$-sphere
is given by
\begin{eqnarray*}
{d{\Omega}_d}^2 = d {{{\theta}_1} ^2} + sin^2 {\theta}_1 d
{{{\theta}_2} ^2}+ sin^2 {\theta}_1 sin^2 {\theta}_2 d
{{{\theta}_3} ^2}\\
+ ...............+\prod_{n=1}^{d-1} sin^2 {\theta}_n
d{{\theta}_d} ^2\\
\end{eqnarray*}
\begin{equation}     \end{equation}
Now, we consider the model which contains a minimally coupled
massless  negative kinetic scalar field. Here the action is taken
as
\begin{equation}  S =  \int d^{d+2}x\sqrt{-g} \left[\frac{R}{2\kappa^2}-
\frac{1}{2}g^{\mu\nu}\partial_{\mu}\phi\partial_{\nu}\phi\right]
\end{equation}
where, $ \kappa^2 = 8 \pi G $. Here the scalar field configuration
is static and spherically symmetric, $\phi_\mu$ can only have one
non-zero component , $\phi_\mu = \phi^{\prime}$. Now, the energy
momentum tensor components in the static wormhole have the form
\begin{equation}  e^{2\beta-2\alpha} T_{tt} = T_{rr} = -\frac{1}{2}(\phi^{\prime})^2    \end{equation}
The equation of motion for $ \phi $,  $ \Box^2 \phi = 0 $ is
\begin{equation}\phi^{\prime \prime} + \left(\alpha^{\prime}-\beta^{\prime} + \frac{d}{r}\right)\phi^{\prime} = 0      \end{equation}
         For the metric (1) and using the above energy momentum tensor , the independent
         Einstein's equations are
\begin{equation}
               \frac{d}{r}\beta^{\prime}+
               \frac{d(d-1)}{2r^2}(e^{2\beta} -1) = - 4\pi G (\phi^{\prime})^2
         \label{Eq3}
          \end{equation}
\begin{equation}
               \frac{d}{r}\alpha^{\prime}-
               \frac{d(d-1)}{2r^2}(e^{2\beta} -1) = - 4\pi G (\phi^{\prime})^2
         \label{Eq3}
          \end{equation}

 \pagebreak

\title{ \underline{\textbf{III. Solutions}}: }

The  field equation (5) for $\phi$ yields
\begin{equation} \phi^{\prime}   = \frac{\phi_0 e^{\beta-\alpha}}{r^d}
\end{equation}
where, $\phi_0$ is an integration constant.

To control   the solutions, we assume the following assumption:

 \begin{equation} \alpha(r) = 0 \end{equation}

\textbf{Argument:} One of the traversability properties is that
the tidal gravitational forces experienced by a traveller must be
reasonably small. So, we assume a zero tidal force as seen by the
stationary observer. Thus one of the traversability conditions is
automatically satisfied.\\
Using equations (8), (9), one gets the solutions from the field
equations (6) and (7) as
\begin{equation}  e^{-2\beta}   =  1 - \frac{b(r)}{r}
\end{equation}
where, the shape function  b(r) is given by
\begin{equation}  b(r)  =  \frac{8 \pi G \phi_0^2}{d(d-1)r^{2d-3}}
\end{equation}
The corresponding scalar field solution is
\begin{equation} \phi  =  \frac{1}{(d-1)\sqrt{\frac{8 \pi G }{d(d-1)}}}
 \cos^{-1}{\sqrt{\frac{8 \pi G \phi_0^2}{d(d-1)r^{2d-2}}}}
\end{equation}
One can note that $\phi  \rightarrow 0$ as $r\rightarrow r_0$
whereas $\phi $ tends to a  constant value when $r >> r_0$.
\\
\\
\\
\\
\\
\\
\\
\\
\title{ \underline{\textbf{IV. Wormhole structure}}: }

For the assumption $\alpha(r) = 0 $ implies no horizon exists in
the spacetime.  Also one can note that $\frac{b(r)}{r}\rightarrow
0 $ as $ \mid r \mid \rightarrow \infty $ i.e. the spacetime is
asymptotically flat. Here the throat occurs at $ r= r_0 $ for
which $ b(r_0) = r_0$.
\begin{equation}  r_0  =  \left[\frac{8 \pi G
\phi_0^2}{d(d-1)}\right]^\frac{1}{{2d-2}}
\end{equation}
 We notice that $b^{\prime}(r_0)< 1$  implies $d>1$. Thus flare-out condition has
been satisfied since our spacetime is greater than or equal to
four dimension. Thus our solution describing a static spherically
symmetric higher dimensional wormhole supported by scalar Fields
with Negative Kinetic Energy.

The axially symmetric embedded surface $ z = z(r)$ shaping the
Wormhole's spatial geometry is a solution of
\begin{equation}
 \frac{dz}{dr}=\pm
 \frac{1}{\sqrt{\displaystyle{\frac{r}{b(r)}}-1}}= \pm
 \frac{1}{\sqrt{\displaystyle{\frac{d(d-1)r^{2d-2}}{8 \pi G \phi_0^2}}-1}}
 \end{equation}
 \\
 \\
  One can note from the definition of Wormhole that at   $ r= r_0 $
  (the wormhole throat) Eq.(14) is divergent i.e.  embedded surface is
   vertical there.

\begin{figure}[htbp]
    \centering
        \includegraphics[scale=.3]{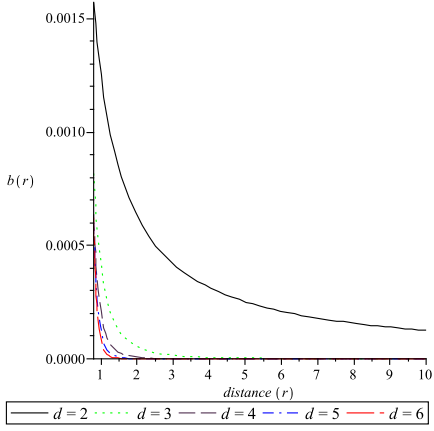}
    \caption{The diagram of the shape function of the wormhole for different dimensions ($\phi_0 = .01,
   G = 1  $ ).   }
    \label{fig:wormhole}
\end{figure}
\begin{figure}[htbp]
\centering
        \includegraphics[scale=.3]{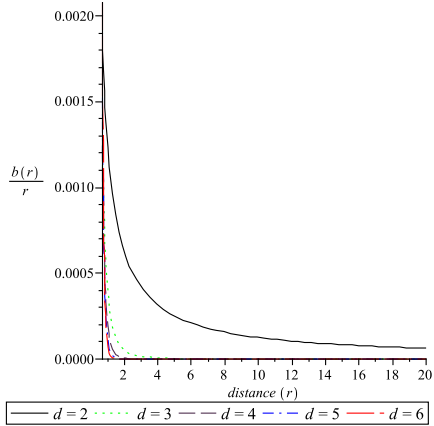}
    \caption{The ratio $\frac{b(r)}{r}$ is plotted for different dimensions ($\phi_0 = .01,
   G = 1  $ ). }
    \label{fig:wormhole}
\end{figure}

\pagebreak

According to Morris and Thorne [3] , the 'r' co-ordinate is
ill-behaved near the throat, but proper radial distance\\
\begin{equation}
 l(r) = \pm \int_{r_0^+}^r \frac{dr}{\sqrt{1-\frac{b(r)}{r}}}
            \label{Eq20}
          \end{equation}
 must be well behaved everywhere i.e. we must require that $ l(r)
 $is finite throughout the space-time . \\
 For our model, one can determine the proper
 distance through the wormhole as ( expanding
binomially in powers of r and assuming $ \phi_0^4$ and higher
powers are zero )
\begin{equation}
 l(r) = \left[ r - \frac{8 \pi G \phi_0^2
 r^{2d-3}}{d(d-1)(2d-3)}\right]
             \label{Eq20}
          \end{equation}
\begin{figure}[htbp]
    \centering
        \includegraphics[scale=.3]{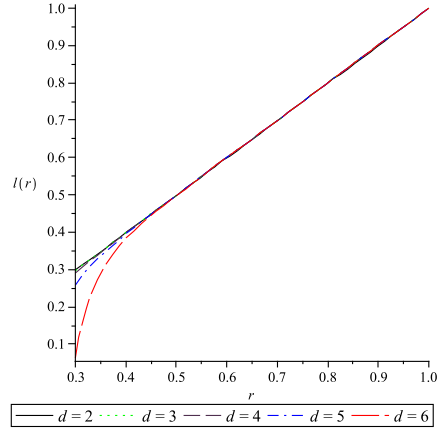}
    \caption{The diagram of the radial proper distance of the wormhole for different dimensions ($\phi_0 = .01,
   G = 1  $ ).  }
    \label{fig:wormhole}
\end{figure}
\begin{figure}[htbp]
    \centering
        \includegraphics[scale=.3]{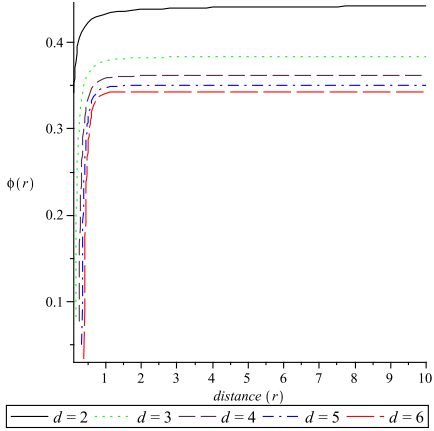}
    \caption{We plot the scalar field $\phi$ for different dimensions ($\phi_0 = .01,
   G = 1  $ ).    }
    \label{fig:wormhole}
\end{figure}
\begin{figure}[htbp]
    \centering
        \includegraphics[scale=.3]{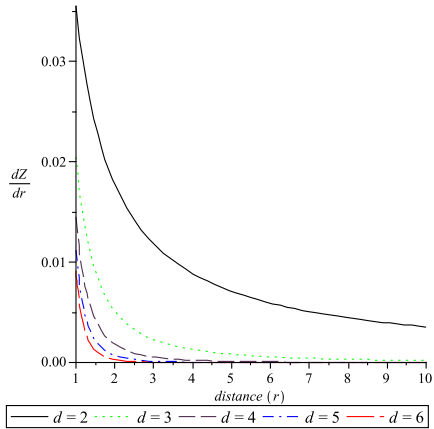}
    \caption{The nature of $\frac{dZ}{dr}$ for different dimensions ($\phi_0 = .01,
   G = 1  $ ). }
    \label{fig:wormhole}
\end{figure}
\\
\pagebreak
\\
\title{ \underline{\textbf{V. Summary  and Discussions}}: }

In this article, we have presented a higher dimensional  wormhole
supported by phantom energy derived from a scalar field with
negative kinetic term. In this model the exotic matter is
described by a scalar field with negative kinetic Energy. The
total amount of average null energy condition (ANEC) violating
exotic matter for this wormhole can be quantified by the integral
[24-25]

 $
            I = \oint ( T_t^t + T_r^r ) dV\\
            =   \int_{r_0}^{\infty}2 \left[ -(\phi^{\prime})^2)e^{-2\beta}
                \frac{2\pi^{\frac{d+1}{2}}}{\Gamma(\frac{d+1}{2})} \right ] r^d
                dr\\
                 = -\frac{4\phi_0^2 \pi^{\frac{d+1}{2}}}{\Gamma(\frac{d+1}{2})(d-1)\sqrt{\frac{8 \pi G
\phi_0^2}{d(d-1)}}}$ .

 Now, we are interested to the fact that
under what conditions the total amount of ANEC violating
 matter could be reduced. One could see that $\phi_0$ and
dimension of the spacetime affect the total amount of ANEC
violating  matter needed. The variation of the total amount of
ANEC violating matter with respect to $\phi_0$ and dimension are
shown in the figures 6-8. If the dimension of the spacetime  is
kept fixed, then the total amount of ANEC violating matter is
reduced by decreasing $\phi_0$ as desired. It is surprising to
note that the total amount of ANEC violating matter needed is
decreasing with respect to dimension greater than eight but up to
dimension eight it is increasing. The figures 6-8 support this
astonishing phenomena. The assumption that the redshift function
to be constant function implies the tidal gravitational force
experienced by a traveller is zero. Thus one of the traversibility
condition is satisfied, in other words, our wormholes containing
small amount of exotic matter, in spite of, they are traversable
for human beings. We have presented an example ( for four
dimensional case )
 of the possible structure of the wormhole  generated by the rotation of the embedded
    curve  about the vertical z axis (see appendix for details).

\begin{figure}[htbp]
    \centering
        \includegraphics[scale=.3]{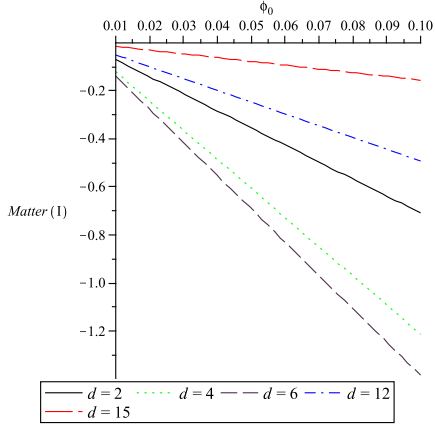}
    \caption{We show the variation of total amount of ANEC violating matter with respect
    to $\phi_0$ for different dimensions ($
   G = 1  $ ).}
    \label{fig:wormhole}
\end{figure}
\begin{figure}[htbp]
    \centering
        \includegraphics[scale=.3]{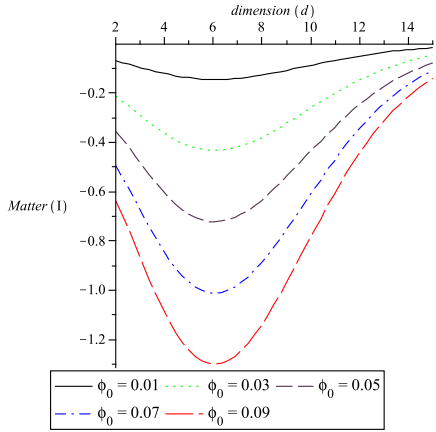}
    \caption{We show the variation of total amount of ANEC violating matter with respect
    to dimension for different  $\phi_0$ ($
   G = 1  $ ).}
    \label{fig:wormhole}
\end{figure}\begin{figure}[htbp]
        \includegraphics[scale=.6]{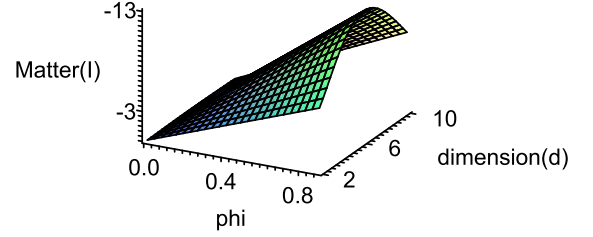}
    \caption{We show the variation of total amount of ANEC violating matter with respect
    to dimension and $\phi_0$ ($
   G = 1  $ ).  }
\end{figure}

\pagebreak

 { \bf \underline{ Acknowledgments:} }

          F.R is thankful to Jadavpur University and UGC , Government of India for providing
          financial support.   \\

\pagebreak

 { \bf \underline{Appendix:}  }

The shape function for the case $d=2$ i.e. for the four
dimensional spacetime is very similar to the case studied by Ellis
[26]. But our work is interesting and  is more justified as we
have taken care of minimizing  the total amount of exotic matter
to be needed to construct the wormhole. In this case, one can get
exact analytical forms of embedded function z and proper radial
distance as \\

$ Z = \pm\sqrt{4 \pi G \phi_0^2} \cosh^{-1}
\frac{r}{\sqrt{4 \pi G \phi_0^2} }$\\

 $ l(r) = \pm \sqrt { r^2 - 4 \pi G \phi_0^2 }$\\

One can see embedding diagram of this wormhole in fig 9. The
surface of revolution of this curve about the vertical z axis
makes the diagram complete (fig10).

\begin{figure}[htbp]
    \centering
        \includegraphics[scale=.3]{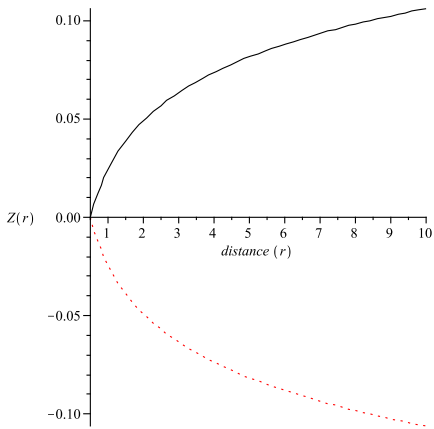}
    \caption{The embedding diagram of the wormhole (  $\phi_0 = .01 $
    and
   $ G = 1  $ ).}
    \label{fig:wormhole}
\end{figure}
\begin{figure}[htbp]
        \includegraphics[scale=.3]{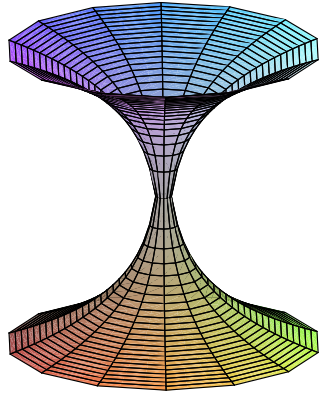}
    \caption{The full visualization of the surface generated by the rotation of the embedded
    curve about the vertical z axis .  }
\end{figure}

\pagebreak

  Due to the simple expression for $l(r)$, one can rewrite the
metric tensor in terms of this proper radial distance,

$
  ds^2 = -  dt^2 + dl^2 + r^2(l) d\Omega_2^2,
$

where

$ r^2(l) = l^2 + 4 \pi G \phi_0^2 $

This is a well behaved coordinate system.

\begin{figure}[htbp]
        \includegraphics[scale=.3]{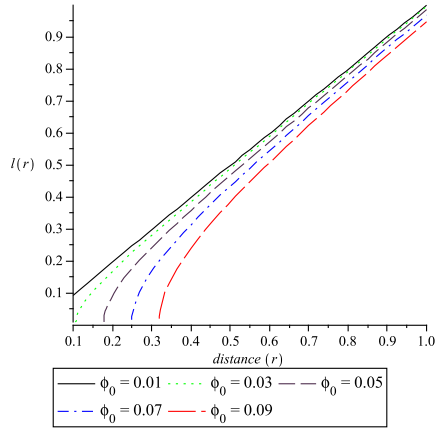}
    \caption{The diagram of the radial proper distance of the wormhole  for different $\phi_0
    $ ( $  G = 1  $ ).  }
\end{figure}

\end{document}